# Quaternary two-dimensional (2D) transition metal dichalcogenides (TMDs) with tunable bandgap

Sandhya Susarla,[†][§] Alex Kutana,[†] [§] Jordan A. Hachtel,[‡] [§] Vidya Kochat,[†] Amey Apte, [†] Robert Vajtai,[†] Juan Carlos Idrobo, [‡] Boris I. Yakobson,[†]* Chandra Sekhar Tiwary,[†]* and Pulickel M Ajayan∗,[†]

[†]Materials Science and Nano Engineering, Rice University, Houston, TX 77005, USA

[‡]Center for Nanophase Materials Sciences, Oak Ridge National Laboratory, Oak Ridge, TN 37831, USA

**Abstract:**

Alloying/doping in two-dimensional material has been important due to wide range band gap tunability. Increasing the number of components would increase the degree of freedom which can provide more flexibility in tuning the band gap and also reduced the growth temperature. Here, we report synthesis of quaternary alloys $Mo_xW_{1-x}S_{2y}Se_{2(1-y)}$ using chemical vapour deposition. The composition of alloys has been tuned by changing the growth temperatures. As a result, we can tune the bandgap which varies from 1.73 eV to 1.84 eV. The detailed theoretical calculation supports the experimental observation and shows a possibility of wide tunability of bandgap.

Corresponding authors: biy@rice.edu (Prof. B. I. Yakobson), cst.iisc@gmail.com (Dr. C. S. Tiwary), ajayan@rice.edu (Prof. P. M. Ajayan)

**Introduction:**

Two-dimensional (2D) transition metal dichalcogenides (TMDs) [1,2,3] have gained considerable importance due to their wide-ranging applications in photonics,[1,2] photodetectors[3,4], optoelectronics[5], electro-catalysis[6,7], energy storage[8] etc. The broad range of TMDs consist of several classes of 2D materials starting from metals ($NbS_2$, $TaS_2$ etc.) to wide bandgap semiconductors ($MoS_2$, $WS_2$ etc.). To tune the bandgap, these materials are often doped/alloyed with other elements[8,9,10], mechanically strained[11], or stacked in the form of heterostructures[12]. Among all these approaches, the doping/alloying provides an easily scalable route to engineer the band structure.[13–15] Exploring the above idea, several efforts have been made in recent past where either anions (transition metal) or cations (chalcogenides) are alloyed with a third element.[16,17,18,19,6,20] Also, theoretical (ab initio) calculations have been performed to predict band gap modification due to alloying. [21–24] Effective carrier masses have been predicted to vary nearly linearly for some bands [24], despite the presence of disorder in the supercell. Use of the cluster expansion method [25] in conjunction with ab initio calculation allows efficient computation of mixing energies and determination of the accurate ground state for structures at different concentrations [13,15,22,24]. In addition to band gap modification, alloying can also result in structural and electronic phase transformation (e.g. 2H to 1T transformation in $Mo_{1-x}W_xTe_2$ alloy systems from parent 2H phase of $MoTe_2$ and 1T' phase of $WTe_2$).[26,27] Alloying can also improve the thermodynamic stability due to increase in entropy. Till now, most of the studies are focused on addition/alloying of third element; further addition of elements provides a larger room to alter the electronic properties. Among TMDs, the sulphides ($MoS_2$, $WS_2$) and selenides ($MoSe_2$ and $WSe_2$) are most explored due to their isostructural nature and band gap variability from 1.6 to 2eV. Combination alloys of both cations and anions have not yet been explored, but can be an interesting ground for exploring thermodynamic stability, microscopic phase segregation to accommodate for charge instability and modification of electronic properties. Such a multicomponent alloy system of atomic thickness with tunable band structure and electronic properties can have compelling prospects in the fields of optoelectronics and catalysis. Here, we present evidence of formation of thermodynamically stable $Mo_xW_{(1-x)}S_{2y}Se_{2(1-y)}$ alloys grown using chemical vapor deposition whose composition is tunable as a function of growth temperature. The structure-property relation of this multicomponent system was mapped out using DFT calculations and further explored using photoluminescence measurements to confirm the predicted variation of electronic band gap. The complex structural arrangement of the individual components of the alloy was investigated using Raman spectroscopy, X-ray photoelectron spectroscopy (XPS) and scanning tunnelling electron microscopy (STEM).

**Results and Discussion**:

In order to explore the bandgaps at different compositions computationally, we generated structures representing random $Mo_{1-x}W_xSe_{2(1-y)}S_{2y}$ alloys with $x=y$. A total of 152 structures were generated, and their geometry optimized at the LDA functional level. The values of the band gap were then extracted from additional calculations where a fine Brillouin zone grid was used. Figure 1a shows schematically the ranges of band gap values of different kinds of alloys. The calculated LDA band gaps for the quaternary $Mo_{1-x}W_xSe_{2(1-x)}S_{2x}$ alloy are shown in Fig. 1b. The alloy displays good gap variability with values spanning the range between 1.60 and 2.03 eV, with a positive bowing parameter. In contrast, band gaps of binary alloys $Mo_{1-x}W_xS_2$ and $MoSe_{2(1-x)}S_{2x}$ [24] allow gap variations in narrower ranges, as also seen from Fig 1b. In all compounds, band gaps at fixed concentrations depend on the atomistic structural details, as evidenced by the multiple values along the ordinate of Fig.1b. In particular, in $Mo_{1-x}W_xSe_{2(1-x)}S_{2x}$, there is largest gap variation at $x=1/2$. We calculated the band structures of $Mo_{1/2}W_{1/2}SeS$ alloys with smallest and largest gaps. The atomic structures for the alloys are shown in Fig. 1c & d, whereas the corresponding band structures are shown as insets in Fig.1b.

Both compounds are seen to be direct-gap semiconductors with band gaps of 1.60 and 1.78 eV, respectively. The two structures have same elemental compositions and similar unit cells, the essential difference between them being site occupancies by the transition metal and chalcogen atoms, as seen in Fig. 1c & d. These alloys thus show great sensitivity of the electronic structure to atomistic details. In structures such as $Mo_{1-x}W_xSe_{2(1-x)}S_{2x}$, where mixing energies are small, one expects presence of different atomic configurations, leading to great local variability of electronic properties. This variability may also have effect on the electronic properties at larger scales

Theoretical calculations were supported by experiments in which we synthesized different compostions of $Mo_{(1-x)}W_xSe_{2(1-y)}S_{2y}$ and observed its effect on the band gap. $Mo_xW_{(1-x)}S_{2y}Se_{2(1-y)}$ were grown using chemical vapor deposition technique (schematic in Fig. 2a, See methods for more details). Maintaining a constant precursor concentration, we have varied the growth temperature for composition tunability. Fig. 2b (i-iv) shows the optical images of the samples grown at 650 $^0$C, 700 $^0$C, 750 $^0$C and 800 $^0$C respectively having varying morphologies. Growth at lower temperature (650$^0$C) resulted in irregular morphology, whereas sharp triangular domains were visible at higher temperatures (above 700 $^0$C). A representative thickness profile and high magnification atomic force microscope (AFM) image of the 2D sheet (as-grown sample at 700 $^0$C) are shown in Fig. 2c (i-ii). The section analysis (Fig. 2 (iii)) shows a thickness of ~0.7 nm thus confirming the thickness to be that of monolayer. AFM of other samples also showed similar thickness thereby confirming them to be monolayer alloys (See Supporting information).

The structural composition of the grown alloys were investigated using Raman spectroscopy of several monolayers (Fig. 3a (i-iv)) as well as spatial mapping of the specific Raman vibrational modes (Fig 3b (i-iv)) . At 650 $^0$C, as-grown sample shows presence of two distinct out-of-plane vibration peaks arising from the $A_{1g}$ modes corresponding to Mo-S and W-S bond vibrations. At higher growth temperatures, these peaks merge into a single peak (Fig. 3a (i)), which is consistent with previous reports of homogeneous alloying without phase separation.[28] The $A_{1g}$ modes of Mo-Se and W-Se bond vibrations (shown in Fig. 3a (iii)), shows a small variation in peak position.[29] We also observe that due to influence of S in the system, Mo-Se and W-Se bond vibrations shift to higher values. For the in-plane vibrations, we observe a three mode behaviour at lower temperatures 650 $^0$C and 700 $^0$C attributing to MoS$_2$-like $E^1_{2g}$, WS$_2$ like $E^1_{2g}$,[28] and MoSe$_2$ like $E^1_{2g}$ [29] vibrations. At higher temperatures 750 $^0$C and 800 $^0$C, these peaks combine into a single broad peak at 329 cm$^{-1}$, indicated by Q, attributing to contributions from Mo-W-S-Se bond vibrations as shown in Fig. 3a (ii). Explanation of evolution of Q peak is described in detail in the supplementary information. In all the alloys grown above 700 $^0$C, we observe a peak at lower wavenumbers (in figure 3a (iv)) which can be attributed to (E2g(S-W)-LA(S-W)+A1g(Se-W)-LA(Se-W)) W-S-Se vibration.[6] Detailed indexing of the individual peaks and their characteristics are discussed in supporting information. .

In order to correlate the morphology with structure, we have obtained Raman maps of individual sheets grown at different temperatures as shown in Fig. 3b (i)-(iv). The four different colours corresponds to intensity variation in a sheet for the wavenumbers: 405 cm$^{-1}$(red), 268 cm$^{-1}$ (cyan), 160 cm$^{-1}$(green) and 325-350 cm$^{-1}$(blue) which corresponds to Mo-W-S, Mo-S-Se, W-S-Se and W-S/W-Mo-S-Se respectively. The 2D sheet grown at 650 $^0$C, shows non-uniform distribution of four phases across the sheet. A high concentration of Mo-W-S, Mo-W-Se (red and green) are observed at edges and also in the form of small clusters distributed across the sheet. The qualitative intensity of W-S-Se (cyan) is lower as compared to all phase. The W-S (350 cm$^{-1}$ peak position, blue) is found to be uniformly distributed. In case of 700 $^0$C grown sheet, the distribution of W-S-Se and W-S are similar to 650 $^0$C grown sheet. But we observe a concentration gradient of Mo-W-S and Mo-W-Se phases from centre to edge. We observe a drastic change for 750 $^0$C and 800 $^0$C grown sheets. In both of these cases, we observe uniform distribution of all four phases across the sheet (Fig. 3(b) (iii)-(iv)).

To develop further insight into the composition and chemical states of the quaternary alloys, X-ray photoelectron spectroscopy (XPS) was performed. A representative XPS spectrum of sample grown at 700 $^0$C is shown in Fig. 3c. The Mo 3d peak (Fig. 3c (i)) splits due to spin-orbit coupling to form separate peaks $3d_{3/2}$ and $3d_{5/2}$ at binding energy of 232.9 eV and 229.8eV respectively indicating $Mo^{4+}$ oxidation state. Similarly, the spin-orbit split W 4f peaks, (Fig. 3c(ii)) exist as $4f_{5/2}$ (at 35.3 eV) and $4f_{7/2}$ (at 33.1 eV) corresponding to $W^{4+}$ oxidation state. In the case of anions, for Se (Fig. 3c (iii)) we observe two peaks corresponding to $3d_{5/2}$ (55.2 eV) and $3d_{3/2}$ (56.1 eV), thus confirming it as $Se^{2-}$ state (selenide). Similarly, for S, we observe two characteristic peaks as $2p_{3/2}$ (162.9 eV) and $2p_{1/2}$ (164.1 eV) which also corresponds to $S^{2-}$ state (sulphide) (Fig. 3c(iv)). We also observe Se $2p_{3/2}$ overlapping in the S 2p region. The alloy composition was calculated using the area under the curve from characteristic peaks which gives the cation to anion ratio as approximately 1:2 which justifies that the as-grown alloys belong to the $MX_2$ class of TMDs. (Detailed analysis of composition and XPS peaks of samples grown at other temperatures are described in the supplementary information.)

To study the intricate atomic distribution of the elements in such a complex quaternary 2D system, the 700 $^0$C as grown samples are examined in a Nion aberration-corrected UltraSTEM 100 operated with an accelerating voltage of 60 kV.[30] Figure 4 (a) shows an atomic-resolution high-angle annular dark field (HAADF) image of the quaternary alloy. In HAADF the primary contrast mechanism is Z-contrast (atomic number) which allows for the atoms to labeled directly from the HAADF intensities.

The HAADF image possesses a large variation in intensities, so to better distinguish between different types of atoms they are first sorted into M sites and $X_2$ sites by examining the coordination of each atom with respect to their nearest neighbors. Due to the hexagonal lattice, all sites should each have three nearest neighbors, but the orientation of the neighbor atoms for the M sites should be flipped compared to the neighbor atoms at the $X_2$ sites: i.e. gone type should have two neighbors above and one below, while the other has two above and one below. Since there are few anti-site defects in the sample this type of sorting is highly effective as can be seen in the $MX_2$ sorted image shown in Figure 4b.

Identifying which site is the metal site and which is the chalcogenide site is done by observing a histogram of the intensities at each of the different sites. The M-site intensities are shown in Figure 4c, while the $X_2$ site intensities are shown in Figure 4d. We know the histogram in 4c shows the metal atoms, and the histogram in 4d shows the chalcogenides because of the effective Z-values of the constituent atoms. The HAADF intensity of an atom goes approximately as $\sim Z^2$ so the order of the predicted relative intensities of the atoms from low to high should be $S_2$ ($Z^2$=256), Mo ($Z^2$=1764), $Se_2$ ($Z^2$=2312), W ($Z^2$=5476). The presence of 5 peaks in the histograms is explained by the possibility of their being chalcogenide columns with both an S and Se atom ($Z^2$=1412).[31] Given the effective Z-contrasts of the different constituent atoms, it is clear the histogram with three peaks is the chalcogenides, and the one with two is the metal sites. Indeed, the peaks of all five columns from the experimental histograms fit extremely well with the predicted order of intensities from the Z-contrasts.

With the peaks identified from the HAADF intensities the type of each atom can be identified to form an atom labeled image shown in Figure 4e. From the image, each element is distributed mostly uniformly throughout the sample, indicating that the elements in the 700 $^0$C have mixed well. The same analysis has also been performed on a larger field-of-view image to show that the mixing is consistent throughout the sample, which is discussed supplementary information.

The variation of phases across the 2D sheet as a function of temperature can be explained with help of thermodynamics and kinetics of growth. In case of growth of individual TMDs, we need a temperature of 750 $^0$C or above but in case of quaternary system we start observing all of these phases at 650 $^0$C growth temperature. In general, the mixing of multi-component to form an alloy results in lowering of melting temperature as compared to the pure component. Due to absence of detailed phase diagram information of the quaternary system, we cannot make quantitative assessment. But multicomponent mixing can be a unique approach to lower growth temperature. Although, we observe presence of all these phases at 650 $^0$C, but their mixing (Q peak in Raman and XPS spectra) starts at 750 $^0$C which can be explained simply by higher mobility of atoms.

To determine the optical properties of the synthesized alloy, photoluminescence measurements were carried out on quaternary alloys as shown in Fig. 5a. The PL emission from 650 $^0$C grown sheet shows two peaks at 1.73 eV and 1.87 eV. It can be attributed to non-uniform mixing of the individual components[2] as observed in structural characterization. As the growth temperature increases, mixing of the four components takes place which results in a single peak (at 1.74 eV) which is again consistent with our structural observations. This single peak position changes from 1.74 eV to 1.84 eV as we vary the growth temperature. Fig. 5b summarizes the PL peaks position of quaternary alloys as compared to individual TMDs. The monolayer of MoSe$_2$, WSe$_2$, MoS$_2$ and WS$_2$ shows an optical band gaps of 1.55 eV, 1.65 eV, 1.85 eV and 1.95 eV respectively. The optical bandgap of quaternary alloys lies in between MoSe$_2$ and WS$_2$. The change in PL peaks with composition in the quaternary alloys can be seen from DFT calculated band bowing diagram in figure 5c. The resulting plot is a surface with positive band bowing parameter. The compositions formed from the experiments have been well fitted on the band bowing surface formed by calculations and are indicated by colored dots on the diagram. The PL mapping was also done on the all the as-grown samples, and is shown in figure 5d (i-iv). PL maps indicate that the optical band gap is uniform for alloys grown at 650 $^0$C but it is highly quenched. At 700$^0$C, the edge and center show different band gaps of 1.75 eV and 1.61 eV respectively due to composition variation from edge to center as indicated by Raman. At even higher temperatures, at 750 $^0$C and 800 $^0$C we observe a uniform band gap all over the flakes with few dark regions corresponding to anion vacancies.

To further understand the optical properties in detail, low temperature measurements were carried out on 700 $^0$C sample as shown in Fig. 5e. We resolve an extra peak at lower energy of 1.42 eV. The previous reports show that this peak arises due to bound excitons created by localized excitons at the defects such as anion vacancies.[34] The variation of intensity and optical band gaps with respect to temperature have been shown in Fig. 5f (i-ii). A blue shift of the main excitonic peak is observed as the temperature increases, typical for most semiconductors. The shift is mostly attributed to dynamic electron-phonon interactions, and to a much lesser extent static lattice dilation.[35] Interestingly, an opposite red shift is observed for the low energy peak. The A-excitonic peak reduces in intensity at lower temperatures indicating bright excitonic character arising due to Mo dominance in the alloy.

In conclusion, with the help of mixing of four elements we can map a wide range of bandgaps as predicted by calculation and further verified using experimental observation. We have successfully synthesized a two-dimensional quaternary alloy (Mo-W-S-Se) using CVD. The composition of alloys has been tuned by varying the temperature from 650 $^0$C to 800 $^0$C. By changing the composition, we could tune the band gap from 1.62 to 1.84 eV, with the position and intensity of the main and low-energy PL peaks showing monotonic temperature variation.

**Methods:**

*Synthesis of quaternary alloys:* Ammonium Molybddate tetrahydrate (AT) (99.999% Sigma Aldrich) and Ammonium Meta Tungstate (AMT) (99.999% Sigma Aldrich) were weighed in equal weight ratio (in the experiment 1.2 mg was used) and kept at different positons in hot zone of the quartz tube as shown in Figure 2a. Selenium(Se) and Sulphur(S) powders were also weighed in equal molar ratio and were mixed properly in mortar and pestle. The obtained mixture was kept in cold zone of the quartz tube kept inside in the tube furnace. Finally, Silicon/Silicon dioxide (Si/SiO$_2$) wafer was kept upside down on AMT powder. This was done because of fact that W has lower vapor pressure as compared to Mo. So, to ensue uniform mixing AMT was kept on Si/SiO$_2$ wafer. On other hand, there is not much difference between melting points of S and Se. So they were kept together. Once the set up was made, samples were grown at four temperature namely 650 $^0$C, 700 $^0$C, 750 $^0$C, 800$^0$C to study the temperature dependence on the growth of these alloys. The ramp rate was 30$^0$C/min and hold time was 20 mins in all the cases

*Characterization:* DFT calculations were performed within the local density approximation (LDA) to the exchange-correlation potential, using the projector augmented wave (PAW) method, as implemented in the Vienna ab initio simulation package (VASP).[36] The kinetic energy cutoff for the plane-wave basis was set to 520 eV in all calculations, and spin-orbit interactions were neglected. The vacuum region was represented by a cell with a size of 15 Å in the *z* direction. System geometries were relaxed until the maximum force on any atom was less than 0.01 eV/Å.

Raman Spectroscopy and Photoluminescence studies (at Room Temperature) were carried out using Renishaw inVia Raman microscopy setup using 532 nm laser with 10% power for Raman spectrum and 1% for PL. The Raman and PL maps were also taken on the same setup step size as 0.5 um and acquisition parameters same as for their respective spectrum. The images were processed using Wire 4.2 software provided by Renishaw. For low temperature PL studies, special low temperature stage by Renishaw was used. Liquid N2 was used as coolant in the set up. X-ray photoelectron measurements were carried out using PHI Quantera XPS with 200 eV Al $K_{alpha}$ X-rays. The XPS peaks were fitted using Multipak software.

**Acknowledgement:**

This work was supported by the MURI ARO program, grant number W911NF-11-1-0362, by FAME, one of six centers of STARnet, a Semiconductor Research Corporation program sponsored by MARCO and DARPA.
**References**

1. Mak, K. F. & Shan, J. Photonics and optoelectronics of 2D semiconductor transition metal dichalcogenides. *Nat Phot.* **10,** 216–226 (2016).

2. Xia, F., Wang, H., Xiao, D., Dubey, M. & Ramasubramaniam, A. Two-dimensional material nanophotonics. *Nat Phot.* **8,** 899–907 (2014).

3. Lopez-sanchez, O., Lembke, D., Kayci, M., Radenovic, A. & Kis, A. Ultrasensitive photodetectors based on monolayer MoS 2. *Nat. Nanotechnol.* **8,** 497–502 (2013).

4. Cheng, R. *et al.* Electroluminescence and Photocurrent Generation from Atomically Sharp WSe2/MoS2 Heterojunction p–n Diodes. *Nano Lett.* **14,** 5590–5597 (2014).

5. Fiori, G. *et al.* Electronics based on two-dimensional materials. *Nat Nano* **9,** 768–779 (2014).

6. Fu, Q. *et al.* Synthesis and Enhanced Electrochemical Catalytic Performance of Monolayer WS2(1-x) Se2x with a Tunable Band Gap. *Adv. Mater.* **2,** 1–7 (2015).

7. Asadi, M. *et al.* Nanostructured transition metal dichalcogenide electrocatalysts for CO2 reduction in ionic liquid. *Science (80-. ).* **353,** 467 LP-470 (2016).

8. Tedstone, A. A., Lewis, D. J. & Brien, P. O. Synthesis , Properties , and Applications of Transition Metal-Doped Layered Transition Metal Dichalcogenides. *Chem. Mater.* **28,** 1965–1974 (2016).

9. Gao, J. *et al.* Transition-Metal Substitution Doping in Synthetic Atomically Thin Semiconductors. *Adv. Mater.* **28,** 9735–9743 (2016).

10. Wang, H., Yuan, H., Hong, S. & Cui, Y. Physical and chemical tuning of two-dimensional transition metal dichalcogenides. *Chem. Soc. Rev.* **4,** 2577–2588 (2015).

11. Johari, P. & Shenoy, V. B. Tuning the Electronic Properties of Semiconducting

Transition Metal Dichalcogenides by Applying Mechanical Strains. *ACS Nano* **6,** 5449–5456 (2012).

12. Hu, X., Kou, L. & Sun, L. Stacking orders induced direct band gap in bilayer MoSe2-WSe2 lateral heterostructures. *Sci. Rep.* **6,** 31122 (2016).

13. Shi, Z., Kutana, A. & Yakobson, B. I. How Much N-Doping Can Graphene Sustain? *J. Phys. Chem. Lett.* **6,** 106–112 (2015).

14. Zeng, Q. *et al.* Band Engineering for Novel Two-Dimensional Atomic Layers. *Small* **11,** 1868–1884 (2015).

15. Shi, Z., Zhang, Z., Kutana, A. & Yakobson, B. I. Predicting Two-Dimensional Silicon Carbide Monolayers. *ACS Nano* **9,** 9802–9809 (2015).

16. Gong, Y. *et al.* Band Gap Engineering and Layer-by-Layer Mapping of Selenium-Doped Molybdenum Disulfide. *Nano Lett.* **14,** 442–449 (2014).

17. Al-Dulaimi, N., Lewis, D. J., Zhong, X. L., Malik, M. A. & O'Brien, P. Chemical vapour deposition of rhenium disulfide and rhenium-doped molybdenum disulfide thin films using single-source precursors. *J. Mater. Chem. C* **4,** 2312–2318 (2016).

18. Suh, J. *et al.* Doping against the native propensity of MoS2: Degenerate hole doping by cation substitution. *Nano Lett.* **14,** 6976–6982 (2014).

19. Xie, L. M. Two-dimensional transition metal dichalcogenide alloys: preparation, characterization and applications. *Nanoscale* **7,** 18392–18401 (2015).

20. Gong, Q. *et al.* Ultrathin MoS 2(1 −x)Se2x Alloy Nanoflakes for Electroctalytic Hydrogen Evolution Reaction. *ACS Catal.* **5,** 2213–2219 (2015).

21. Komsa, H.-P. & Krasheninnikov, A. V. Two-Dimensional Transition Metal Dichalcogenide Alloys: Stability and Electronic Properties. *J. Phys. Chem. Lett.* **3,** 3652–3656 (2012).

22. Kang, J., Tongay, S., Li, J. & Wu, J. Monolayer semiconducting transition metal dichalcogenide alloys: Stability and band bowing. *J. Appl. Phys.* **113,** 143703 (2013).

23. Xi, J., Zhao, T., Wang, D. & Shuai, Z. Tunable Electronic Properties of Two-Dimensional Transition Metal Dichalcogenide Alloys: A First-Principles Prediction. *J. Phys. Chem. Lett.* **5,** 285–291 (2014).

24. Kutana, A., Penev, E. S. & Yakobson, B. I. Engineering electronic properties of layered transition-metal dichalcogenide compounds through alloying. *Nanoscale* **6,** 5820–5825 (2014).

25. Sanchez, J. M., Ducastelle, F. & Gratias, D. Generalized cluster description of multicomponent systems. *Phys. A Stat. Mech. its Appl.* **128,** 334–350 (1984).

26. Voiry, D. & Chhowalla, M. Chem Soc Rev Phase engineering of transition metal dichalcogenides. *Chem. Soc. Rev.* **44,** 2702–2712 (2015).

27. Duerloo, K.-A. N. & Reed, E. J. Structural Phase Transitions by Design in Monolayer Alloys. *ACS Nano* **10,** 289–297 (2016).

28. Song, J. *et al.* Controllable synthesis of molybdenum tungsten disulfide alloy for vertically composition-controlled multilayer. *Nat. Commun.* **6,** 1–10 (2015).

29. Zhang, M. *et al.* Two-Dimensional Molybdenum Tungsten Diselenide Alloys : and Electrical Transport. *ACS Nano* **8,** 7130–7137 (2014).

30. Krivanek, O. L. *et al.* An electron microscope for the aberration-corrected era. *Ultramicroscopy* **108,** 179–195 (2008).

31. Pennycook, S. J. in *Scanning Transmission Electron Microscopy: Imaging and Analysis* (eds. Pennycook, S. J. & Nellist, P. D.) 1–90 (Springer New York, 2011). doi:10.1007/978-1-4419-7200-2_1

32. Wienold, J., Jentoft, R. E. & Ressler, T. Structural Investigation of the thermal decomposition pof ammonium heptamolybdate by in-situ XAFS and XRD. *Eur. J. Inorg. Chem.* 1058–1071 (2003).

33. Hunyadi, D., Sajó, I. & Szilágyi, I. M. Structure and thermal decomposition of ammonium metatungstate. *J. Therm. Anal. Calorim.* **116,** 329–337 (2014).

34. Tongay, S. *et al.* Defects activated photoluminescence in two-dimensional semiconductors: interplay between bound, charged, and free excitons. *Sci. Rep.* **3,** 2657 (2013).

35. Manoogian, A. & Woolley, J. C. Temperature dependence of the energy gap in semiconductors. *Can. J. Phys.* **62,** 285–287 (1984).

36. Kresse, G. & Furthmüller, J. Efficient iterative schemes for ab initio total-energy calculations using a plane-wave basis set. *Phys. Rev. B* **54,** 11169--11186 (1996).

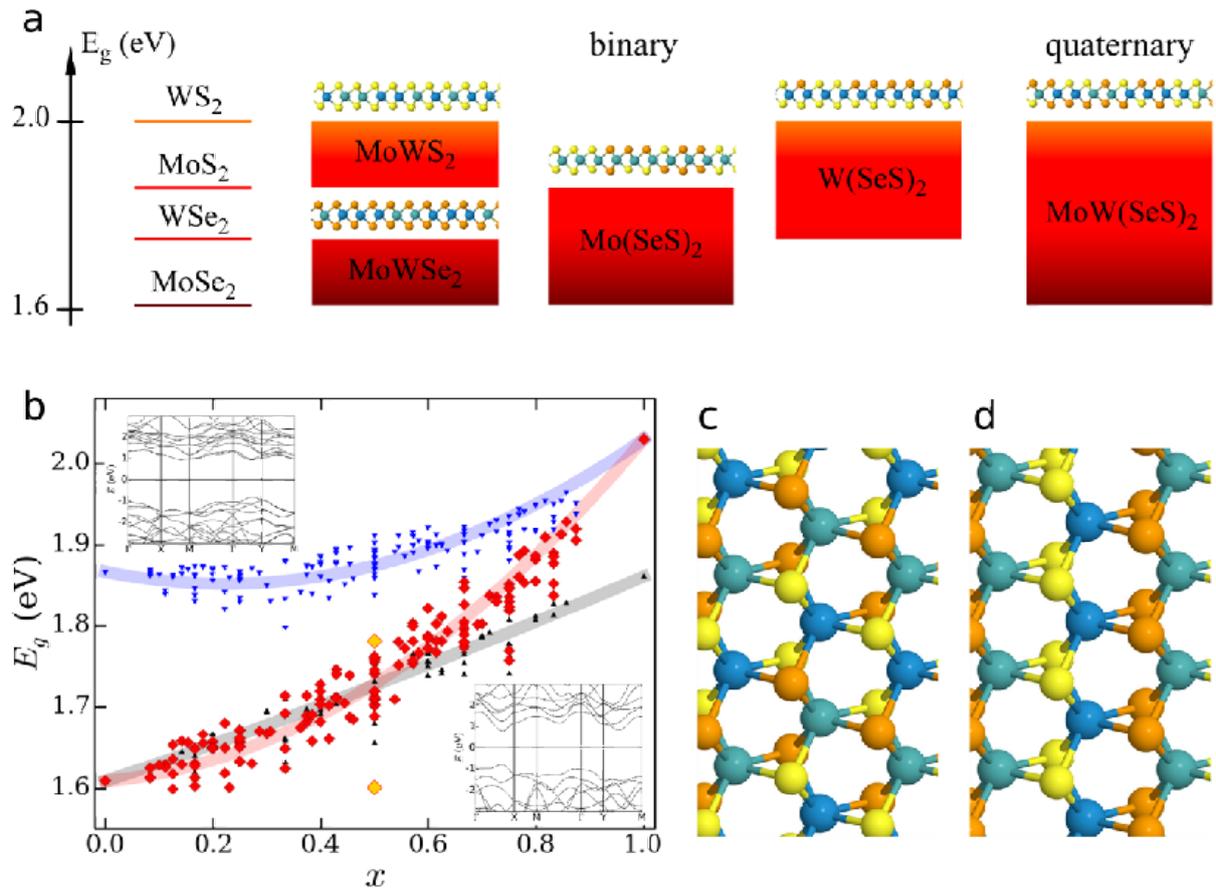

Fig.1: (a) Band gap tuning in MoW(SeS)$_2$ alloys. (b) Calculated band gaps of quaternary Mo$_{1-x}$W$_x$Se$_{2(1-x)}$S$_{2x}$ alloy (large red diamonds), and binary Mo$_{1-x}$W$_x$S$_2$ and MoSe$_{2(1-x)}$S$_{2x}$ alloys (small blue downward and black upward triangles, respectively). Band structures of two Mo$_{1-x}$W$_x$Se$_{2(1-x)}$S$_{2x}$ alloys at $x$=1/2 with largest and smallest band gaps are shown as insets at top left and bottom right, respectively. The band gap values for the two structures are highlighted with orange diamonds and atomic structures are shown in (c) and (d). Both compounds are direct-gap semiconductors with band gaps of 1.78 and 1.60 eV, respectively.

.

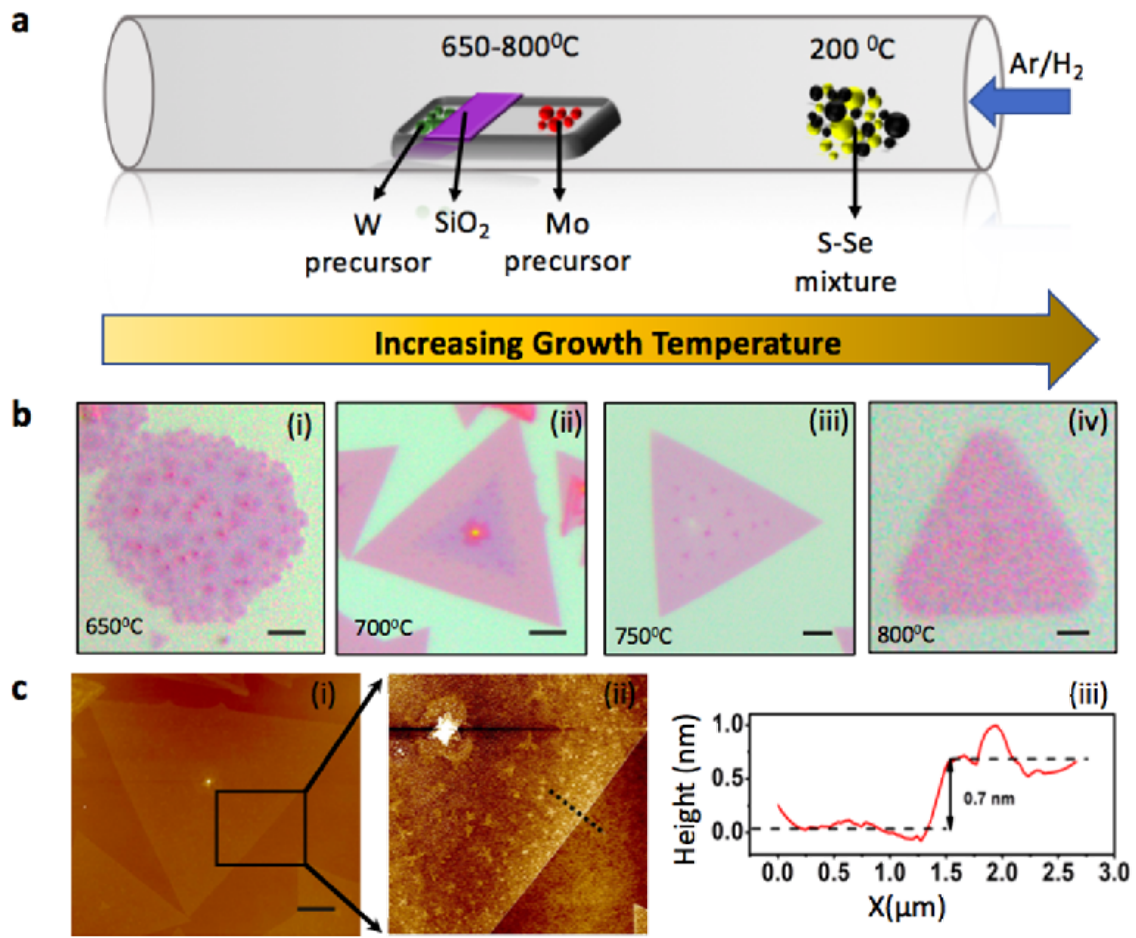

Fig.2 **a)** Schematics of growth procedure for quaternary alloys. b) Optical Images of monolayer quaternary alloys grown at i) 650°C ii) 700°C iii) 750°C iv) 800°C. c) (i-ii) AFM images and iii) height profile of monolayer grown at 700°C Scale bars: b (i-iii), c(i): 5 μm (iv) 2 μm

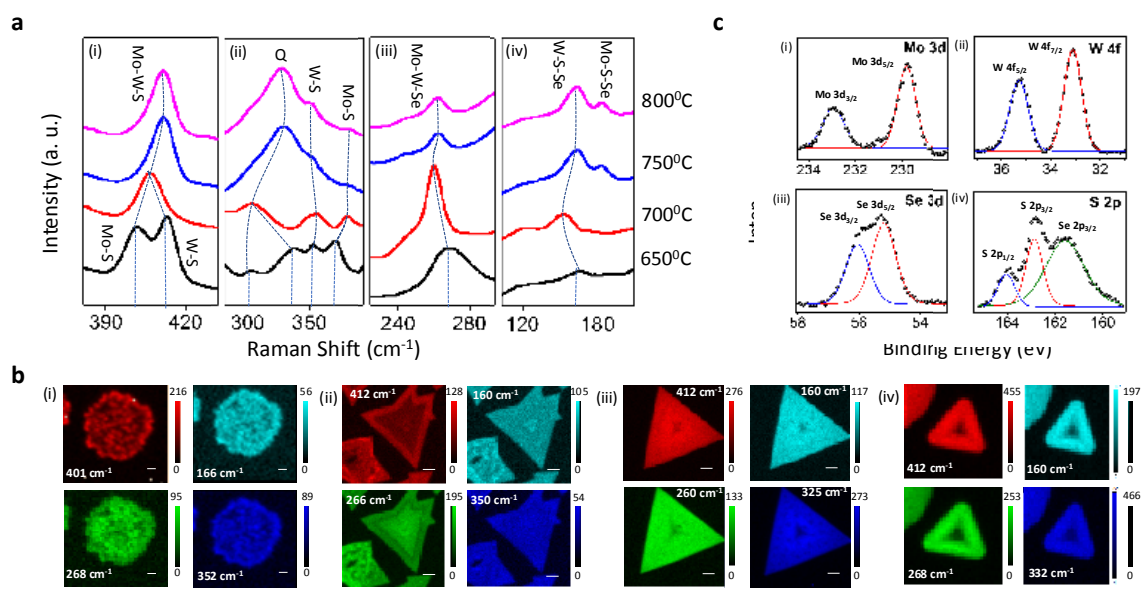

Fig.3 a) (i-iv) Raman spectroscopy of as grown alloy at different temperatures. The dotted lines indicate the trend of merging of various vibrational peaks at different growth temperature b) Raman mapping of alloys grown at i: 650°C ii: 700°C iii: 750°C iv: 800°C showing the presence of Mo-W-S (red), Mo-W-Se (green), W-S-Se (cyan), W-S/W-Mo-S-Se(blue). c) X-ray photoelectron spectroscopy showing binding states of : i) Mo 3d ii) W 4f iii) S 2p iv) Se 3d Scale bars: i,iv: 2μm ii,iii: 5 μm

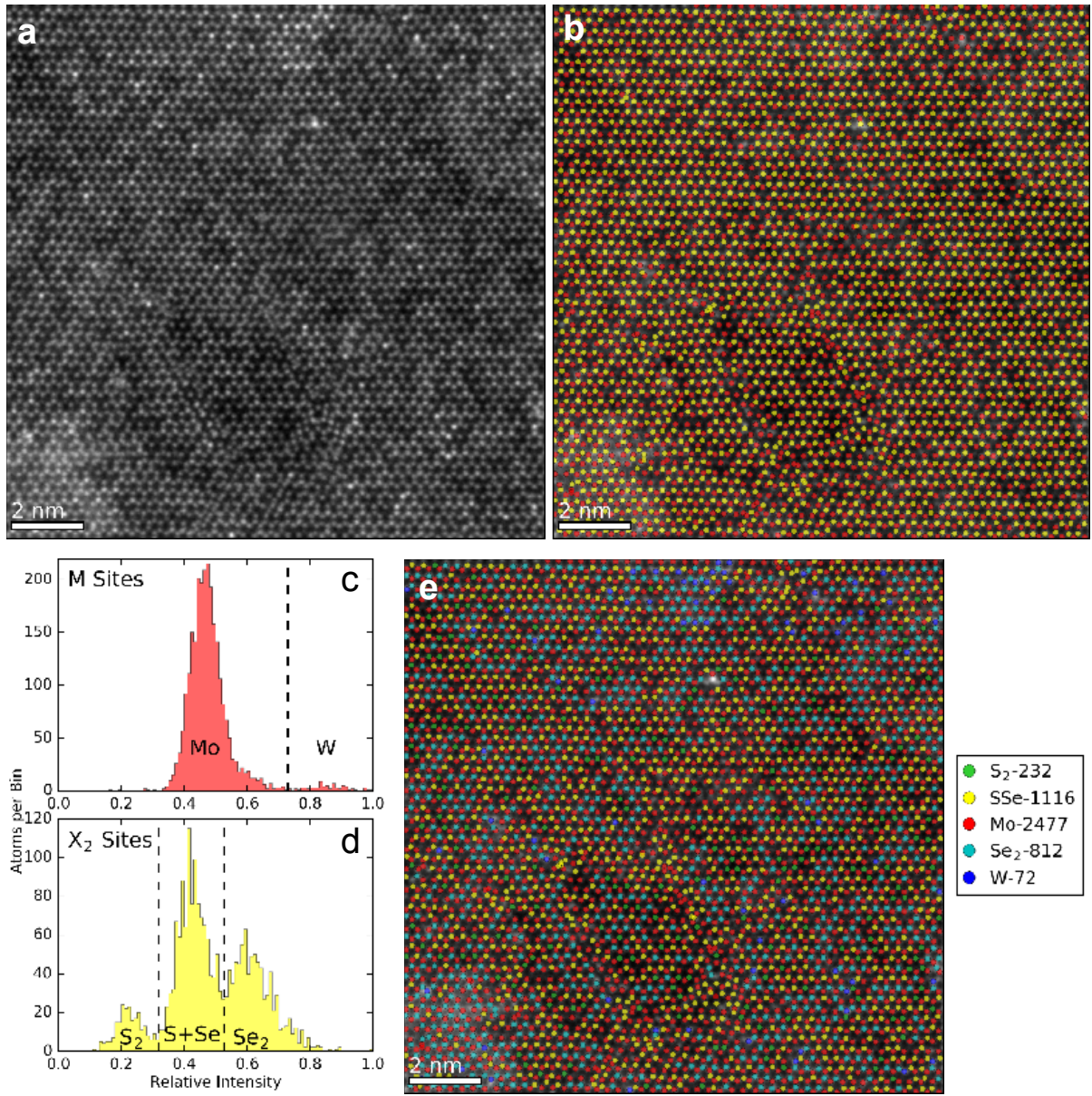

Fig.4: a) High resolution STEM-HAADF image of monolayer as grown sample at $700^0$C. b) Atoms in the image are found and sorted into metal (M) sites and chalcogenide ($X_2$) sites based off their atomic coordination. c-d) The HAADF intensity histograms of the M and $X_2$ sites respectively. The M histogram shows two distinct peaks while the $X_2$ histogram shows three peaks, the peaks are then labeled as Mo and W in the M sites, and $S_2$, S+Se, and $Se_2$ sites based on the relative Z-contrasts. e) The histograms in c) and d) are used to label all the atoms in b) to create a final atom-labeled image of the sample, showing that the constituent atoms uniformly alloyed throughout the material.

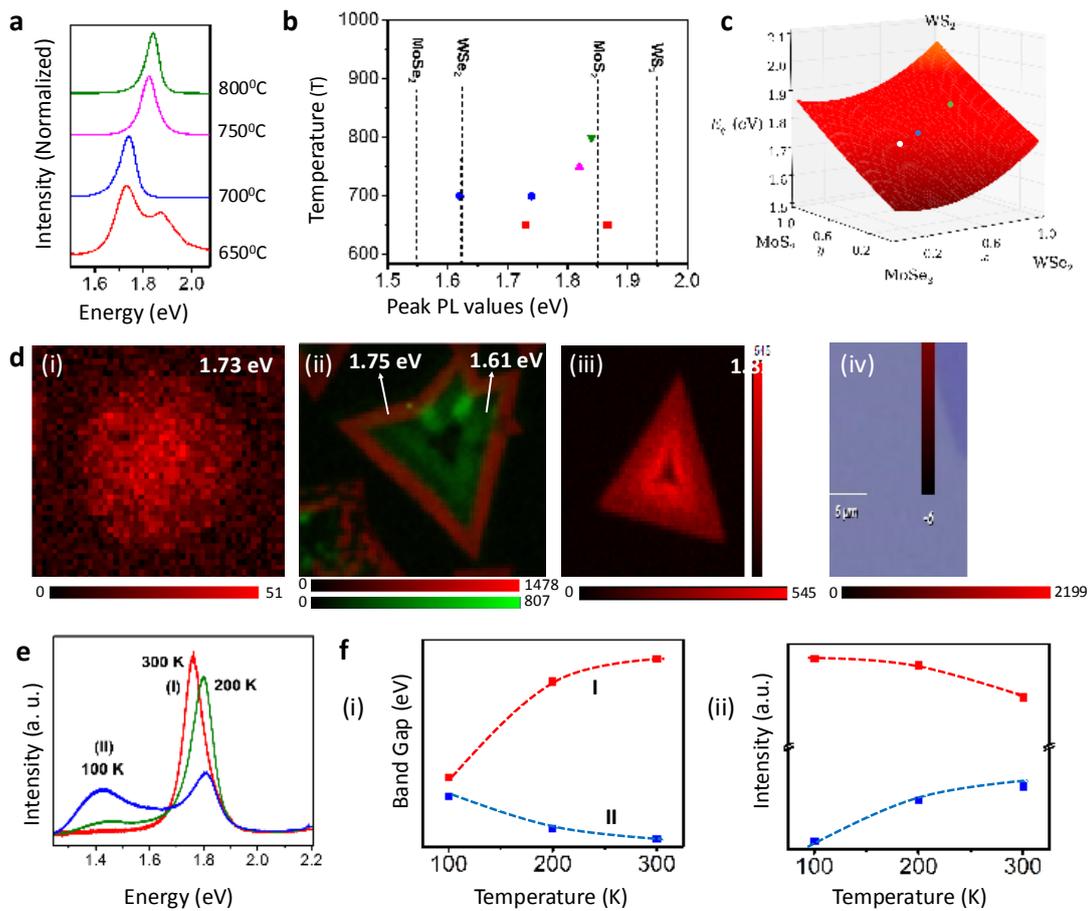

Fig.5: a) Room temperature photoluminescence (PL) spectra of as grown samples. b) Comparative plot showing PL peak positions of quaternary samples as compared to pure 2-D materials like $MoS_2$, $WS_2$, $MoSe_2$ and $WS_2$ c) 3-D band bowing diagram showing the variation of peak positions with composition. Experimental composition points have been indicated as dots as white (650$^0$C), blue (700$^0$C) and green (750$^0$C). d) PL maps of as grown alloys at i) 650$^0$C ii) 700$^0$C iii) 750 $^0$C iv) 800$^0$C e) Low temperature PL of the as grown sample at 700$^0$C at three different temperatures (100-300 K) showing two prominent peaks namely I and II. f) plots showing the variation of i) intensity ii) position of PL peaks with temperature.